# Feasibility Enhancement of Constrained Receding Horizon Control Using Generalized Control Barrier Function

Haitong Ma
School of Vehicle and Mobility
Tsinghua University
Beijing, China
maht19@mails.tsinghua.edu.cn

Xiangteng Zhang
School of Aerospace Engineering
Tsinghua University
Beijing, China
zhangxt18@mails.tsinghua.edu.cn

Shengbo Eben Li*
School of Vehicle and Mobility
Tsinghua University
Beijing, China
lishbo@tsinghua.edu.cn

Ziyu Lin
School of Vehicle and Mobility
Tsinghua University
Beijing, China
linzy17@mails.tsinghua.edu.cn

Yao Lyu
School of Vehicle and Mobility
Tsinghua University
Beijing, China
lyo.tobias@foxmail.com

Sifa Zheng
School of Vehicle and Mobility
Tsinghua University
Beijing, China
zsf@tsinghua.edu.cn

*Abstract*—Receding horizon control (RHC) is a popular procedure to deal with optimal control problems. Due to the existence of state constraints, optimization-based RHC often suffers the notorious issue of infeasibility, which strongly shrinks the region of controllable state. This paper proposes a generalized control barrier function (CBF) to enlarge the feasible region of constrained RHC with only a one-step constraint on the prediction horizon. This design can reduce the constrained steps by penalizing the tendency to move towards the constraint boundary. Additionally, generalized CBF is able to handle high-order equality or inequality constraints through extending the constrained step to nonadjacent nodes. We apply this technique on an automated vehicle control task. The results show that compared to multi-step pointwise constraints, generalized CBF can effectively avoid the infeasibility issue in a larger partition of the state space, and the computing efficiency is also improved by 14%-23%.

## I. Introduction

Receding horizon control(RHC) is a common technique to handle optimal control problem (OCP) with state constraints. A typical RHC procedure is that at each time step the OCP is solved over a finite horizon, and only the first control element is implemented. The notorious infeasibility issue of RHC often happens due to the existence of state constraints, which means that there may exist a feasible solution at the initial time step, but fails to obtain a feasible solution in a certain subsequent time step, even with a perfect model. Infeasible issue is strongly related to the feasible region where no infeasibility issue happens. Feasible region is usually a subset of constrained set, and the explicit formulation with general dynamic systems can hardly be obtained.

A straightforward approach to handling state constraints is to adopt the real-world safety constraints on each step of the prediction horizon [1][2]. It only guarantees that state constraints are not violated in the current prediction horizon, while feasibility usually requires that state constraints are obeyed in not only current but also all subsequent prediction horizons. Extending the length of prediction horizon is helpful to improve feasibility, but the horizon length selection usually relies on experience. Therefore, theoretical guarantee is pursued for feasibility consideration. Early studies about feasibility in RHC borrow from ideas of stability enhancement in the late 1990s, which add constraint or cost penalty on the terminal step of prediction horizon [3]. Parisini et al. (1997) integrates terminal cost with constraints that the terminal states must be in a specific set to further improve feasibility [4]. Terminal constraint or cost methods still pose requirements on length of prediction horizon. Some studies aim to find an explicit feasible region. For linear systems, the constrained region is a polyhedron [1]. A feasible region recognition method for linear system uses an iterative algorithm which converges to several polyhedral partitions, but it is only appropriate for time-invariant, linear and low-order systems [5].

The feasibility of RHC with general systems is studied in the Lyapunov-like methods, i.e., control barrier function. Control barrier function (CBF) is a 'dual' of control Lyapunov function for feasibility, and they share the same idea for confining states inside an invariant set. Blanchini (1999) firstly introduces that the feasible region in control theory is an invariant set [6]. Continuous-time CBF is proposed as an inequality constraint on constraint function derivatives lying only on the boundary of constrained set [7]. Another branch of CBF called barrier certificate emerges at the same time, which is initially not relevant with controlled invariance [8]. Barrier certificate indicates that all states of the unsafe set should satisfy the monotonous conditions of constraint function [9]. Compared to the state-of-the-art CBF, barrier certificate is more conservative which renders every subset of feasible region to be invariant [10]. Continuous-time CBF is extended by Ames et al. (2017) to an inequality constraint applicable for all safe states. Conditions inside the set are relaxed compared to the boundary, which renders only the safe set to be invariant [11]. Discrete-time CBF is also discussed and applied, for instance, in a bipedal robot navigation task by Agrawal et al. (2017) [12].

Recent studies provide that CBF can be applied in the online control task to pursue safe exploration by restricting the incoming action due to a more conservative constraint [13]. However, existing studies on using CBF in RHC procedure still adopt CBF on each step of the prediction horizon [14][15]. This design is so redundant that the exact condition of controlled invariance is repeated unnecessarily. Moreover, an implicit drawback of current discrete-time CBF is that the first-order derivative of constraint function must be relevant

This study is supported by National Key R&D Program of China with 2020YFB1600200. This study is also supported by Tsinghua University-Toyota Joint Research Center for AI Technology of Automated Vehicle. All correspondences should be sent to S. E. Li.

with input in the continuous case. Existing studies for high-order, or high relative-degree constraints only focus on continuous-time systems. Nguyen et al. (2016) discusses the continuous-time high relative-degree constraint, and augments new states by Lie derivatives [16]. There is a lack of research on the high-order constraints with the discrete-time dynamic systems, which is common in numerical implementation of OCPs. In conclusion, despite various existing methods listed above considering infeasibility issues with CBF and other methods, they all suffer from some problems: a) Existing feasibility guarantee of RHC relies on length selection of prediction horizons. A bad length selection will result in shrinking of the feasible region. b) Existing discrete-time CBF integrated with RHC procedure is not able to handle high-order constraints, and also repeated redundantly.

Therefore, this paper proposes a discrete-time generalized control barrier function to solve abovementioned issues. The main contributions of this paper can be summarized:

(1) The proposed generalized CBF constraint is able to enlarge feasible region of RHC with only a one-step constraint in the prediction horizon. It suppresses the tendency towards constraint boundary for the purpose of replacing commonly used multi-step constraints, which releases both the computational burden and requirement that existing feasibility enhancement techniques rely heavily on length selection of prediction horizon.

(2) The design of single-step constraint face challenges on high-order constraint. We also improve the discrete-time CBF by posing constraints on nonadjacent steps to handle high-order constraints for wider applicability of the single-step constraint design. This design is able to avoid infeasible issue in a larger partition of state space. Addtionally, the computing time decreases for 14%-23%.

The rest of this paper is organized as follows. Section II is the preliminaries about infeasibility issue and feasible region. Section III introduces the separating domain mechanism of receding horizon control. We introduce a single CBF constraint, and its generalization on high-order constraints. Section IV provides the verification with an automated vehicle. Section V concludes the paper.

## II. PRELIMINARIES

### A. State Constraints and Infeasibility Issue

Ideally, in receding horizon control, we can construct the infinite-horizon optimization problem to guarantee solid feasibility, but an infinite-horizon optimization is usually hard to solve. In practice, we solve a finite-horizon alternative instead. Consider a general state constraint in the discrete-time formulation

$$h(x_{t+i}) \leq 0, i \in \{1,2,\ldots N\} \quad (1)$$

where $x \in \mathbb{R}^n$ is the state, $h: \mathbb{R}^n \to \mathbb{R}$ is the constraint function. A discrete-time time-invariant dynamical system is

$$x_{t+1} = f(x_t, u_t) \quad (2)$$

where $u \in \mathbb{R}^m$ is the control input, and $f: \mathbb{R}^n \to \mathbb{R}^n$ is the environment dynamics. We assume that $f$ is Lipschitz continuous on a compact set containing the origin, i.e., there exists a continuous policy so that the system is asymptotically stable. Moreover, the exact model $f$ is known.

The utility function is denoted as $l(x_{t+i}, u_{t+i}): \mathbb{R}^n \times \mathbb{R}^m \to \mathbb{R}$. The finite-horizon optimization problem becomes

$$J_{RHC} = \sum_{i=0}^{N-1} l(x_{t+i}, u_{t+i})$$

with

$$u_t^* = u_{t+i}^* = \arg \min_{u_{t+i}, i \in \{0,1,2,\cdots,N-1\}} J_{RHC} \Big|_{i=0} \quad (3)$$

s.t.

$$x_{t+i+1} = f(x_{t+i}, u_{t+i})$$
$$h(x_{t+i}) \leq 0, i \in \{1, \ldots N\}$$

The main drawback with finite-horizon optimization is the infeasibility issue. Intuitively, adopting shorter prediction horizon will deprive the ability to predict longer and push the policy to be more aggressive, which will fail to get a feasible solution of the following OCPs. Without loss of generality, a 4-step example is given in Fig. 1. The OCP in the virtual time domain has solutions at $t^{\text{th}}$ and $(t+1)^{\text{th}}$ step, but no solution exists on the $(t+2)^{\text{th}}$ step, i.e., whatever the action is taken, the state trajectory must break the constraints in the $(t+2)^{\text{th}}$ prediction horizon.

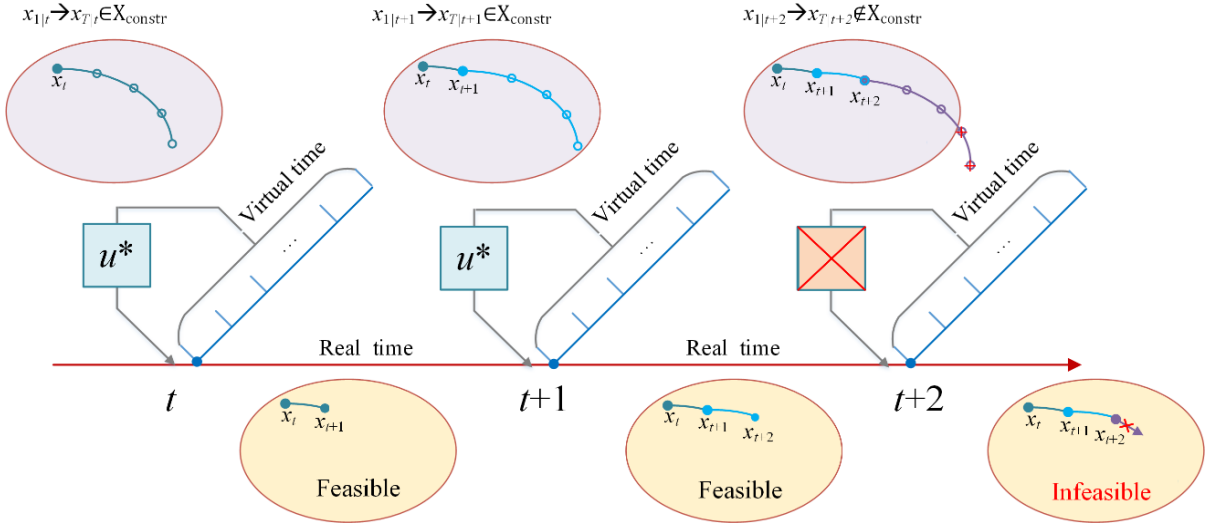

Fig. 1. Demonstration of infeasibility issue.

Trajectory starting from state $x_t$ could have always obtained feasible solution in real time domain, but ends in infeasibility issue due to a shorter prediction of state constraints in the virtual domain. This is the case we want to avoid as much as possible, i.e., we want to design the state constraints properly to avoid infeasibility issue and then enlarge feasible region.

### B. Discrete-time Control Barrier Function

Control barrier function (CBF) poses constraints based on current constraints function value. Existing CBF confines the state in the next step as

**Theorem 1** (Discrete-time control barrier function) The sublevel set $\{x|h(x) \leq 0\}$ is controlled forward invariant along the trajectories if and only if there exists a policy satisfying

$$\big(h(x_{t+1}) - h(x_t)\big) + \lambda h(x_t) \leq 0 \quad (4)$$

with a scalar $\lambda \in (0,1)$ for all states in the sublevel set, and $h(\cdot)$ is called the control barrier function.

**Proof** $h(x_{t+i}) \leq (1-\lambda)^i h(x_t) \leq 0$, $\forall i \in \mathbb{Z}_+$. A detailed proof is presented in [12].

All existing studies combining CBF with RHC procedure directly adopt CBF on each step of the prediction horizon like (5), which is similar to the multi-step pointwise constraint. We use multi-step CBF constraints to denote it and the comparison is shown in Fig. 3.

$$h(x_{t+i+1}) \leq (1-\lambda)h(x_{t+i}), i \in \{0,1,2,\dots N-1\} \quad (5)$$

## III. GENERALIZED CONTROL BARRIER FUNCTION

### A. Reduce to One Inequality Constraint

The RHC procedure can be regarded as that, at $t^{\text{th}}$ step, the optimization problem in the virtual time domain is solved, and the first control input is used to drive the system in the real time domain. We define the separating time domain mechanism to further distinguish the optimization problem and control implementation in the formulation of virtual and real time domain, as shown in Fig. 2.

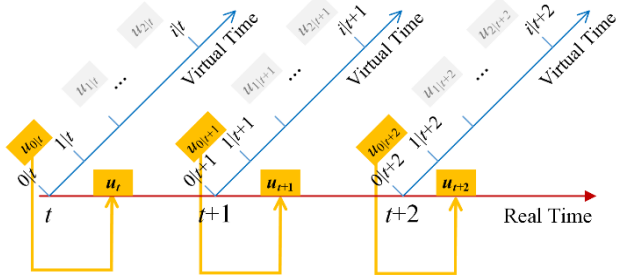

Fig. 2. Control implementation in RHC procedure.

We can conclude that avoiding infeasibility issue requires only the implemented control to be properly constrained. Therefore, by assuming the corresponding subset is controlled invariant, we can only put a single-step CBF constraint in the prediction horizon using the notion in Fig. 2 like

$$h(x_{1|t}) \leq (1-\lambda)h(x_{0|t}) \quad (6)$$

which is different to (5) as shown in Fig. 3. The hyper-parameter $\lambda$ adjusts the conservativeness of policy, which plays a similar role of adjusting constrained steps in traditional multi-step pointwise constraints. Detailed proof of feasibility guarantee will be given later with GCBF constraint.

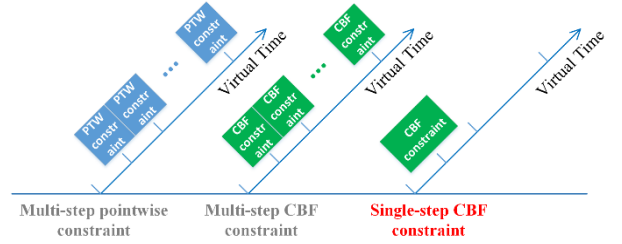

Fig. 3. Single step CBF constraint.

### B. Generalized Control Barrier Function

Although single-step CBF constraint is able to handle infeasibility issue, it implicitly requires that the constraint function on the next step must be affected by the control input. We give a simple example that the condition usually does not hold. Consider an emergency breaking scenario as shown in Fig. 4, states consist of both distance and speed, and the input is defined as desired acceleration, which is a second-order derivative of distance. The discrete-time state-space equation is

$$x' = \begin{bmatrix} d' \\ u' \end{bmatrix} = \begin{bmatrix} 1 & -T \\ 0 & 1 \end{bmatrix}\begin{bmatrix} d \\ u \end{bmatrix} + \begin{bmatrix} 0 \\ T \end{bmatrix}a \quad (7)$$

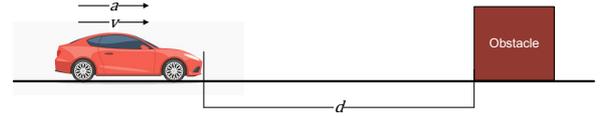

Fig. 4. Emergency braking scenario.

It is easy to know that the constraint is the distance must be greater than 0. The first-step constraint function is

$$h(x_{1|t}) = [1 \quad -T]x_{0|t} + 0 * u_{0|t} \quad (8)$$

where the first-step constraint function is irrelevant with the implemented control input. There will be no constraint on the implemented control with single-step CBF constraint.

Constraint function with similar properties like this distance constraint, whose first order derivate does not contains control input, is called high-order constraints. Existing discrete-time CBF is not able handle high-order constraints, but the problems do not emerge in existing multi-step design of CBF constraints with RHC procedure. It does raise challenges when we reduce the multi-step CBF constraint to a single-step CBF constraint. To overcome this problem of single-step CBF, we propose the generalized control barrier function (GCBF) for high-order constraints. The discrete-time system actually flattens the high-order constraint to time steps, so we just need to pose constraint on a future step. The comparison of GCBF constraint with mentioned constraints is shown in Fig. 5.

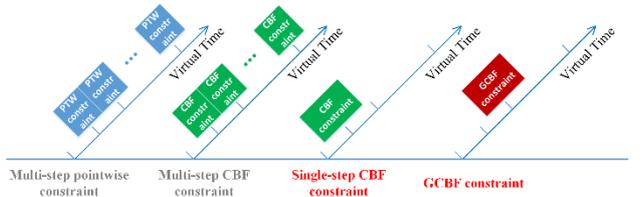

Fig. 5. Comparison of GCBF constraint with other constraints.

Firstly, we define the relative-degree to represent this relation between constraint and input:

**Definition 1** The constraint has relative-degree $m$ with respect to control input if

$$\frac{dh(x_{i|t})}{du_{0|t}} = 0 \tag{9}$$

for $\forall i \in \{0,1,\ldots m-1\}, \forall x \in \mathbb{R}^n$, with respect to system (2), $m \in \{1,2,\ldots n\}$. High-order constraints are those with relative-degree greater than 1.

Existing discrete-time CBF can be regarded as the constraints must have relative-degree 1. We give the generalized control barrier function for constraints with relative-degree greater than 1:

**Theorem 2** For a relative-degree $m$ state constraints $h(x) \leq 0$ and the corresponding sublevel set, assume the system satisfies $h(x_{t+i}) \leq 0$ for all $i \in \{1,2,\ldots m-1\}$, the corresponding sublevel set is controlled forward invariant along the trajectories if and only if there exists a policy satisfying

$$h(x_{t+m}) \leq (1-\lambda)^m h(x_t) \tag{10}$$

with a scalar $\lambda \in (0,1]$ for all states.

**Proof** For $i \geq m$, $h(x_{t+i}) \leq (1-\lambda)^{i-s} h(x_{t+s}) \leq 0$, $\forall i \in \mathbb{Z}_+$, and $s \equiv i \bmod m$, i.e., the remainder of $i$ divided by $m$.

Moreover, we can derive the corollary that the feasibility guarantee exists with GCBF constraint in RHC procedure if the constrained set is controlled invariant.

**Corollary 1** Assume there exists the forward invariant safe set $\{x|h(x) \leq 0\}$, then there exists a scalar $\lambda \in (0,1]$ and solution $u_i^* = u_{0|i}^*$, $i \in \{0,1,\ldots \infty\}$ which will guarantee the state to be confined in the safe set.

**Proof** Consider a state trajectory $\{x_0, x_1, \ldots x_t, \ldots x_\infty\}$ is generated from RHC procedure with GCBF constraint, where $x_0$ belongs to the controlled forward invariant set $\mathcal{C}$. To retain the state trajectory inside $\mathcal{C}$ is equivalent to that every element of the implemented control sequence $\{u_0, u_1, \ldots, u_\infty\}$ should satisfy the conditions of GCBF (10). Note that the control element is equal to $\{u_{0|0}, u_{0|1}, \ldots, u_{0|\infty}\}$ constrained by

$$h(x_{m|t}) - (1-\lambda)^m h(x_{0|t}) \leq 0 \tag{11}$$

Since the exact model in known, (11) is exactly the GCBF condition in Theorem 2. The set invariance is proved.

**Remark 1** The key of the GCBF constraint is that it does not need to constrain the following steps in virtual time domain, and only confining the first control element is enough for feasibility consideration in RHC procedure as shown in Fig. 6. The assumption of the existence of controlled invariant set *usually does not hold*, but the GCBF constraint still achieves remarkable effectiveness on avoiding infeasibility issue by enlarging feasible region.

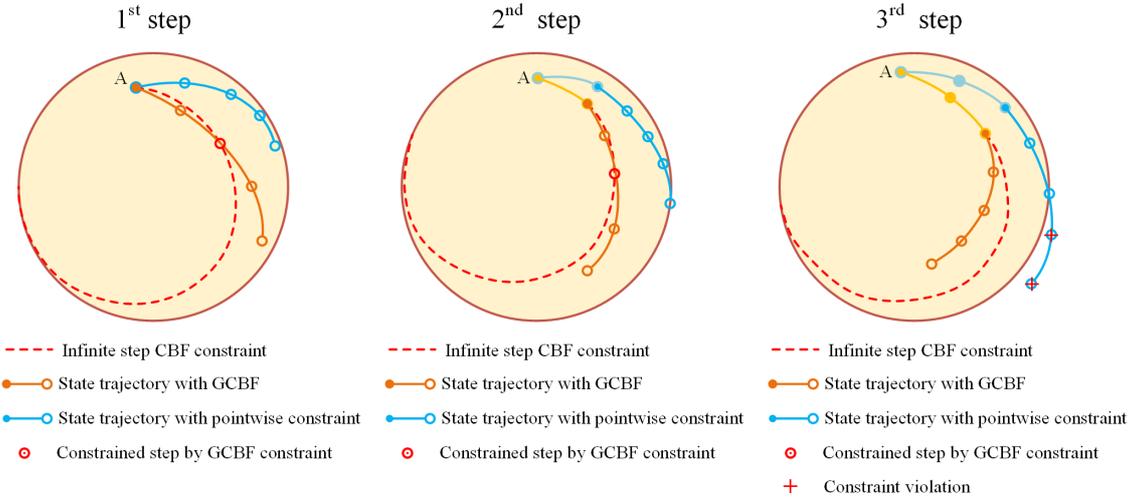

Fig. 6. GCBF constraints to avoid infeasibility issue.

## IV. Experiments

In this section, we evaluate our GCBF constraint on an adaptive cruising control task to show the ability to enlarge feasible region and decrease computing burden. Adaptive cruising control is a safety-critical vehicle longitudinal control task with time-variant nonlinear model[17][18]. Different environmental variants are set like dynamic constraints and unknown disturbance, compared with pointwise constraints.

### A. Problem Discription

The adaptive cruising control (ACC) adopts the three-dimensional vehicle lateral dynamics system whose states and input are listed in TABLE I.

TABLE I. STATE AND CONTROL INPUT

| | | | |
|---|---|---|---|
| State | Distance error with desired distance | $\Delta d$ | [m] |
| | Speed error with proceeding vehicle | $\Delta v$ | [m/s] |
| | Vehicle acceleration | $a_f$ | [m/s$^2$] |
| Input | Desired acceleration | $a_{fdes}$ | [m/s$^2$] |

The states are calculated by

$$\begin{aligned} \Delta d &= d - d_{des} \\ \Delta v &= v_p - v_f \end{aligned} \tag{12}$$

where $d$ is the distance between the cars, and $d_{des}$ represents the desired car distance. The speed of the preceding vehicle and controlled vehicle are denoted as $v_p$ and $v_f$, and $a_f$ describes the acceleration of the controlled vehicle. The desired distance of ACC is described as a quadratic clearance(QDC) model with Taylor expansion

$$d_{des} = av_f^2 + bv_f + c$$
$$= rv_f(v_f - v_{fmean}) + \tau_h v_f + d_0 \quad (13)$$

The relation between vehicle acceleration and desired acceleration is simplified to a first order inertia system, i.e., a generalized vehicular longitudinal dynamics(GLVD) model

$$a_f = \frac{K_G}{T_G s + 1} a_{fdes} \quad (14)$$

Combine (13) and (14) with Newton mechanics, the discrete state-space model of ACC is expressed as

$$x' = f(x,u)$$
$$= \begin{pmatrix} 1 & T & -\tau_h T - rT(2v_f - v_{fmean}) \\ 0 & 1 & -T \\ 0 & 0 & 1 - \frac{T}{T_G} \end{pmatrix} x$$
$$+ \begin{pmatrix} 0 \\ 0 \\ K_G T/T_G \end{pmatrix} u + \begin{pmatrix} 0 \\ T \\ 0 \end{pmatrix} q \quad (15)$$

where $q = a_p$ is the acceleration of proceeding vehicle, which is an observable disturbance[19]. The vehicle parameters are listed in TABLE II.

TABLE II. VEHICLE MODEL PARAMETERS

| Coefficient of drivers' behavior | $r$ | 0.054 | [s$^2$/m] |
|---|---|---|---|
| Time headway | $\tau_h$ | 1.0 | [s] |
| Stop distance | $d_0$ | 2.9 | [m] |
| Step time | $T$ | 0.1 | [s] |
| Gain of GVLD model | $K_G$ | 1.05 | [-] |
| Time constant of GVLD model | $T_G$ | 0.393 | [s] |

The utility function is

$$l(x,u) = 0.02\Delta d^2 + 0.025\Delta v^2 + 5a_{fdes}^2 \quad (16)$$

The input is bounded as $a_{fdes} \in [a_{fmin}, a_{fmax}]$. Considering unexpected actions of proceeding car, the state constraint is defined as [20]

$$\Delta d + d_{des} \geq d_{s0} + \text{TTC} \cdot \Delta v \quad (17)$$

Substitute the QDC model into constraint, we get the state constraint with respect to states

$$h(x) = (r(v_p - \Delta v - v_{fmean}) + \tau_h - \text{TTC})\Delta v$$
$$-\Delta d - \tau_h v_p + d_{s0} - d_0 \leq 0 \quad (18)$$

The parameters in state and input constraints are listed in TABLE III.

TABLE III. CONSTRAINT PARAMETERS

| Minimum acceleration | $a_{fmin}$ | 5 | [m/s$^2$] |
|---|---|---|---|
| Maximum acceleration | $a_{fmax}$ | -5 | [m/s$^2$] |
| Safe distance | $d_{s0}$ | 5 | [m] |
| Time to collision | TTC | -2.5 | [s] |

*B. Algorithm Details and Results*

The original infinite-time OCP is transferred to a finite-horizon OCP in virtual time domain as

$$\min_{u_{t+i}, i \in \{0,1,\ldots\infty\}} \sum_{i=0}^{N-1} l(x_{i|t}, u_{i|t})$$

s.t.

$$x_{i+1|t} = f(x_{i|t}, u_{i|t}) \quad (19)$$

*with pointwise constraints*

$$h(x_{i|t}) \leq 0, i \in \{1,2,\ldots,N_c\}$$

*with barrier constraints*

$$h(x_{m|t}) \leq (1-\lambda)^m h(x_{0|t})$$

Firstly, we compare the OCPs' performance with barrier constraint and pointwise constraints. All optimal control problems use the prediction horizon $N = 50$. The conservativeness hyper-parameter of GCBF is set as $\lambda = 0.01$. The results are shown in Fig. 7, where PTW is short for pointwise constraints.

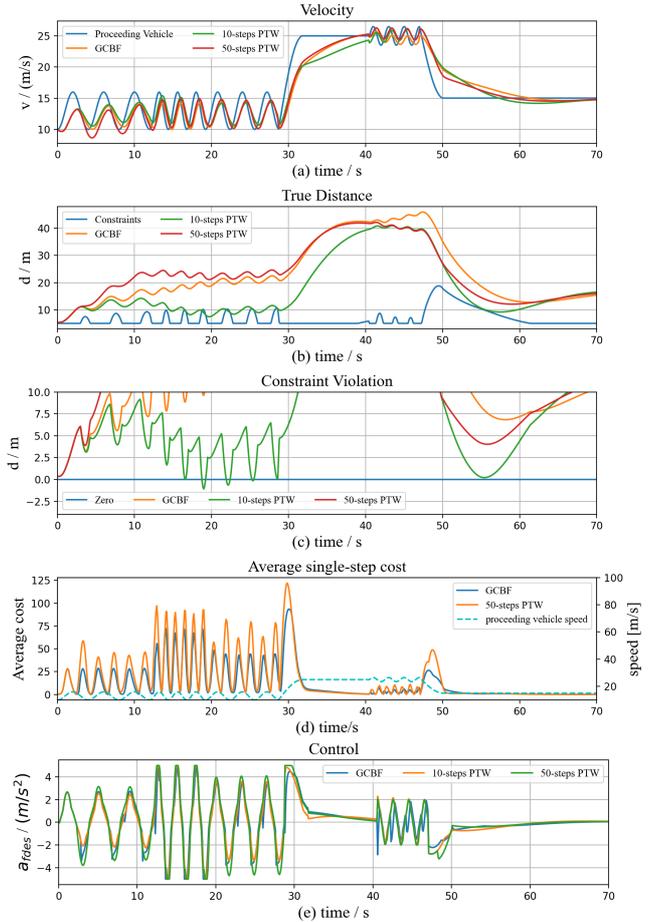

Fig. 7. Performance comparison between GCBF and pointwise constraints.

The design of constraints affects the feasibility directly. Fig. 7(a) and (b) show the speed and the true distance between two vehicles, and Fig. 7(c) demonstrates the difference between true distance and distance constraints. Results show that the 50-steps pointwise constraints and GCBF constraint guarantee that no constraint violation happens, while it is another story with 10-steps pointwise constraints. Fig. 7(e) also depicts that the 10-steps pointwise constraints is likely to take insufficiently large acceleration or deceleration so it suffers from constraints violation.

Constraints also have influence on optimality. The average costs of GCBF and 50-step pointwise constraints are shown in Fig. 7(d). The 50-step pointwise constraints suffer larger cost when the preceding vehicle speed fluctuates quickly. One reason is that 50-steps is too conservative to obtain optimal solution. The other reason, which is more critical in this case, is from the prediction error. The solution of constrained optimization relies heavily on the active constraints, and constraints design is relevant to acceleration of proceeding vehicle, which is assumed to be invariant in the whole virtual time domain. The active constraint of GCBF is always on the second step, where the assumption holds convincingly. On the contrary, the active constraints of 50-steps pointwise constraints design mostly lie on the further steps, for instance, the 40$^{th}$ step in prediction horizon, where the assumption mostly fails in this case.

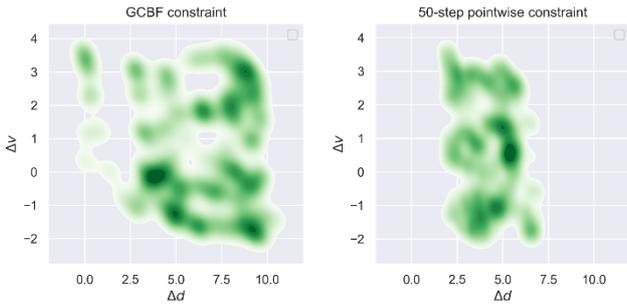

Fig. 8. Visual demonstraintion of feasible region.

Moreover, we give a visual demonstration of enlarging feasible region of GCBF. The state distribution shown in Fig. 8 consists of all feasible states of ten trajectories starting at different initial states. For convenience, the three-dimensional state space is projected to the two-dimensional subspace, which the dimension of acceleration is compressed for more intuitive demonstration. It is easy to see that GCBF obtains a slightly larger feasible region, which lies on the left side of the figure where the ego vehicle is closer to proceeding vehicle. The feasible region distribution with GCBF lies also on the right side of the figure, which means that GCBF tends to keep a further distance than 50-step pointwise constraints to avoid infeasibility issue.

Another significant advantage of GCBF is that simplifying the constraints formulation improves the computational efficiency. We examine the computing time using three optimization packages with different nonlinear programming techniques, including ipopt, Bonmin, and sequential quadratic programming (SQP). Computing time of GCBF and 50-step pointwise constraints is compared in Fig. 9, and details are shown in TABLE IV. Obvious reductions of computing time are achieved since the numbers of inequality constraints decreases. Both interior-point line-search and sequential quadratic programming algorithms get benefits in computing time with decreasing the number of inequality constraints.

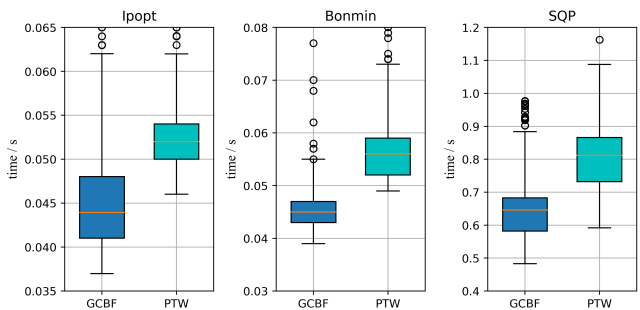

Fig. 9. Calculation time under different state constraint designs.

TABLE IV. AVERAGE SINGLE STEP COMPUTATION TIME

| Optimization package | Average computation time [ms/step] | | Reduction |
|---|---|---|---|
| | GCBF | Pointwise | |
| Ipopt | 45.76 | 53.49 | 14.46% |
| Bonmin | 45.10 | 58.73 | 23.21% |
| SQP | 644.46 | 806.95 | 20.14% |

## V. CONCLUSION

This paper proposes the generalized control barrier function to enlarge feasible region with only a single-step constraint on the prediction horizon. We also extend the discretized control barrier function to high relative-degree state constraints by posing constraints on nonadjacent steps with respect to the relative degree. Moreover, we remove the redundancy that each step in the prediction horizon needs a CBF constraint, and only utilize a one-step GCBF constraint to enhance feasibility and decrease computing burden. We apply our constraints design on an automated vehicle control task. Visual demonstration shows that GCBF avoids infeasibility issues among larger partition of the state space, and reduction of computing time is verified with different optimization techniques.

This study provides a promising approach for handling and simplifying safety-critical receding horizon control problems, various constraints formulation can be redesigned based on this research for avoiding infeasibility issues.

ACKNOWLEDGMENT

We would like to acknowledge Prof. Jianyu Chen, Dr. Jingliang Duan, Mr. Guanya Shi, and Mr. Yao Mu for their valuable suggestions throughout this research.